\documentclass[aps,superscriptaddress, showpacs,pra]{revtex4}

\begin{document}
\title{Rotational invariance as an additional constraint on local realism}

\author{Tomasz Paterek}
\affiliation{Instytut Fizyki Teoretycznej i Astrofizyki,
Uniwersytet Gda\'nski, PL-80-952 Gda\'nsk, Poland}

\author{Wies{\l}aw Laskowski}
\affiliation{Instytut Fizyki Teoretycznej i Astrofizyki,
Uniwersytet Gda\'nski, PL-80-952 Gda\'nsk, Poland}

\author{Marek {\. Z}ukowski}
\affiliation{Instytut Fizyki Teoretycznej i Astrofizyki,
Uniwersytet Gda\'nski, PL-80-952 Gda\'nsk, Poland}
\affiliation{Institut f\"ur Experimentalphysik, Universit\"at
Wien, Boltzmanngasse 5, A--1090 Wien, Austria}
\affiliation{Tsinghua University, Beijing, China}


\title{On Series of Multiqubit Bell's Inequalities}

\begin{abstract}
We overview series of multiqubit Bell's inequalities which apply to
correlation functions. We present conditions that quantum states must satisfy 
to violate such inequalities.
\end{abstract}

\date{\today}

\maketitle

\section{Introduction}

Quantum mechanics gives predictions in form of probabilities.
Already some of the fathers of the theory were puzzled with question
whether there can exist
a deterministic structure beyond quantum mechanics which recovers quantum statistics
as averages over ``hidden variables'' (HV).
In his famous impossibility proof Bell made precise assumptions about the form of 
possible underlying HV structure that allows
mutually distant systems to be independent of one another \cite{BELL64}.
He derived the inequality which must be satisfied by all such (local realistic) structures
and presented the example of quantum predictions which violate it.
In this way the famous Einstein-Podolsky-Rosen (EPR) paradox was solved \cite{EPR}.

Noncommutativity of quantum theory precludes simultaneous deterministic predictions 
of measurement outcomes of complementary observables.
For the EPR this indicated that  
``the wave function does not provide a complete description
of physical reality''.
They expected the complete theory 
to predict all possible measurement outcomes,
prior to and independent of the measurement (realism),
and not to allow
``spooky action at a distance'' (locality).
Such a completion was disqualified by Bell.

A more general version of two-particle Bell's theorem was given 
by Clauser, Horne, Shimony, and Holt (CHSH), and still extended by Clauser and Horne (CH) \cite{CHSH,CH}.
The important feature of the CHSH and CH inequalities,
which hold for all local realistic theories,
is that they can be not only compared with ideal quantum predictions,
but also with experimental results.
Thus a debate that seemed quasi-philosophical, could be tested in the lab
(see \cite{CS} for review of early experiments).

The three or more particle, or as we now say qubit, versions of Bell's
theorem were presented by Greenberger, Horne, and Zeilinger (GHZ),
surprisingly $25$ years after the original Bell's paper \cite{GHZ}.
In contradistinction with the two particle case,
now the contradiction between local realism and quantum mechanics
could be shown for perfect correlations.
Immediately after that Mermin produced series of inequalities for arbitrary many particles,
which cover the GHZ case, and made the GHZ paradox directly testable
in the laboratory \cite{MERMIN}.
A complementary series of inequalities was introduced by Ardehali \cite{ARDEHALI}.
In the next step Belinskii and Klyshko gave series of two settings inequalities, 
which contained the tight inequalities of Mermin and Ardehali \cite{BELINSKII}.
Finally the full set of tight two setting per observer $N$-party 
Bell inequalities for dichotomic observables 
was found independently in Refs. \cite{WW,WZ,ZB}. All these series of inequalities
are a generalization of the CHSH ones \cite{CHSH} .
Such inequalities involve only $N$ party correlation functions. 

The process of finding new series of Bell inequalities 
(for many qubits, many settings per observer, involving all possible correlations
- for arbitrary experimental arrangement)
continues until this day, and is the topic of this overview.
We shall try to present the current state-of-the-art,
concentrating on multiqubit Bell inequalities for correlation functions.
We will \emph{not} give a detailed analysis of the assumptions behind Bell's inequalities.
The reader can find them in excellent papers like \cite{CS,GHSZ}.

With the emergence of the new sub-branch of physics (and information theory),
Quantum Information, the Bell theorem, and Bell inequalities found
applications far away from foundations of quantum physics.
The security analysis of the first entanglement based quantum cryptography
scheme involves Bell's inequalities \cite{EKERT}.
This now is strengthen by the analysis of Scarani and Gisin,
who showed that violation of Bell inequality is indeed
a valid security criterion in quantum crypto-key distribution \cite{SCARANIGISIN}.
Recently it was shown that with every Bell inequality one can associate
a specific quantum communication complexity problem.
With the use of quantum states which violate such Bell inequality
one can always construct a quantum communication complexity problem
which outperforms all possible classical ones \cite{BZPZ}.
This result, as well as the one of Scarani and Gisin,
as well as many other ones, suggest that violation of Bell
inequality is a criterion of \emph{direct} usefulness
of entanglement in quantum information processing \cite{USEFUL_ENT}.
The authors of this overview think that there are very many open questions 
associated with future generalisations of series
of multiqubit Bell inequalities,
and that it is still a fascinating field of studies,
which will find new applications, and new surprising results.

\section{Bell and Clauser-Horne-Shimony-Holt}

The whole history of the EPR paradox
the reader can find in the beautiful review by Clauser and Shimony \cite{CS}.
Twenty nine years after the EPR paper
Bell proved that the completion of quantum mechanics 
expected by EPR, is impossible \cite{BELL64}.
In his original proof Bell utilized the perfect anticorrelations which arise
whenever Alice and Bob measure local spins (with respect to the same direction)
on the two-qubit system in the state:
\begin{equation}
|\psi^- \rangle = \frac{1}{\sqrt{2}}\Big[ |0 \rangle_A | 1 \rangle_B - |1 \rangle_A |0\rangle_B \Big],
\label{SINGLET}
\end{equation}
where $|0\rangle$ and $|1\rangle$ denotes the eigenbasis of the local $\sigma_z$ operator.
However unavoidable experimental imperfections imply that 
correlations are never perfect.
Here we re-derive the CHSH inequality,
for which perfect correlations do not have to be assumed
and thus the inequality can be directly
experimentally checked.
The violation of CHSH inequality implies that no local realistic
explanation for the observed correlations is possible.
But what if the inequality is satisfied?
Can we then build a local realistic model for our observations?
The answer is negative.
Necessary and sufficient condition for local realistic model
involves a set of inequalities, not a single one.

The pair emission begins an experimental run.
Alice and Bob measure one of two alternative settings each run.
Their choices what to measure are absolutely free,
uncorrelated (statistically independent) with the workings of the source.
According to realism all possible measurement outcomes
exist prior to and independent of the measurement.
Moreover locality assumes that the outcomes of Alice depend on her setting only,
and the same for Bob.
We denote their predetermined local realistic results as $A_1$, $A_2$ for Alice,
and $B_1$, $B_2$ for Bob.
For example if Alice chooses to measure setting ``1''
she obtains the outcome $A_1$, if she chooses to measure ``2''
she obtains $A_2$.
Under realism assumption
all possible measurement outcomes are defined,
even if only some of them are actually measured.
The experiments on qubits can give one of two results, to which
we ascribe numbers, $+1$ and $-1$, i.e. $A_k,B_l = \pm 1$.
We form a ``vector'' out of the predetermined results of each observer:
$\vec A = (A_1,A_2)$ and $\vec B = (B_1,B_2)$ in this case.  
One can also define a ``vector'' 
(or if you like, a ``tensor'')
of local realistic correlation functions,
$\hat E_{LR}$, with components $E^{LR}(k,l) = \langle A_k B_l \rangle_{\text{avg}}$,
where the average is taken over many experimental runs.
Then all such local realistic models, $\hat E_{LR}$, can be written as:
\begin{equation}
\hat E_{LR} = \sum_{\vec A, \vec B = (\pm1,\pm1)} P(\vec A,\vec B) \vec A \otimes \vec B,
\label{CORRELATION_POLYTOPE}
\end{equation}
where $P(\vec A,\vec B)$ is the probability with which a certain quadruple of predetermined results
$\{A_1,A_2,B_1,B_2\}$ appears.
That is, every local realistic model of the correlation functions
is a convex combination of the extreme points $\vec A \otimes \vec B$,
and thus lies within a convex polytope, spanned by the vertices $\vec A \otimes \vec B$.
The necessary and sufficient condition for local realistic description
is a set of inequalities which define the interior of the polytope
and are saturated at the border hyperplane of it.
Such inequalities are called \emph{tight} Bell's inequalities.

Let us present a construction of the necessary and sufficient condition for 
the possibility of local realistic description of correlation functions for such 
experiments with two qubits.
First we derive a necessary condition for local realistic model,
then construct such a model proving that the condition is also sufficient.
We introduce a more elaborated notation for future use in the generalisation
to arbitrary number of particles.
The two local dichotomic observables are 
parametrised by vectors $\vec{n}_1^j$ and $\vec{n}_2^j$, 
and the predetermined results for the $j$th party by
$A_j(\vec{n}_1^j) = \pm 1 $ and $A_j(\vec{n}_2^j) = \pm 1$,
as for now $j=1,2$ ($1$ for Alice, $2$ for Bob).
Since $A_j(\vec{n}_{k}^j)=\pm 1$, for each observer
$j$ one has either $|A_j(\vec{n}_1^j)+A_j(\vec{n}_2^j)|=0$ and
$|A_j(\vec{n}_1^j)-A_j(\vec{n}_2^j)|=2$, or vice versa.
Therefore, for all sign choices of $s_1,s_2=\pm 1$ the product
$[A_1(\vec{n}_{1}^1) + s_1 A_1(\vec{n}_{2}^1)][A_2(\vec{n}_{1}^2) + s_2 A_2(\vec{n}_{2}^2)]$ vanishes
except for one sign choice, for which it is equal to $\pm 4$.
If one adds up all such four products, with an arbitrary sign in
front of each of them, the sum is always  equal to the value of
the only non-vanishing term, i.e., it is $\pm 4$.
Thus the following algebraic identity holds for the predetermined results:
\begin{equation}
A_{12,12;S} \equiv \sum_{s_1,s_2 =\pm1} 
S(s_1,s_2)[ A_1(\vec{n}_{1}^1) + s_1 A_1(\vec{n}_{2}^1)]
[A_2(\vec{n}_{1}^2) + s_2 A_2(\vec{n}_{2}^2)] =\pm 4, \label{INEQ2}
\end{equation}
where $S(s_1,s_2)$ stands for an arbitrary ``sign'' function of the
summation indices $s_1,s_2$, such that its values
are only $\pm1$. 
The notation $A_{12,12;S}$ describes the situation where two parties
choose between two settings ``1'' or ``2''.

After averaging the expression (\ref{INEQ2}) over the ensemble of
the runs of the experiment one
obtains the following set of Bell inequalities:
\begin{equation}
\Big|
 \sum_{s_1,s_2 = -1,1} S(s_1,s_2) 
 \sum_{k_1,k_2 = 1,2}
 s^{k_1-1}_1 s^{k_2-1}_2 E({k_1},{k_2}) \Big| \leq
4. \label{allbellineq}
\end{equation}
Since there are 16 different functions $S(s_1,s_2)$,
the inequalities (\ref{allbellineq}) represent a set of $16$
Bell inequalities for the correlation functions. 
A specific choice of the sign function,
$S(s_1,s_2) = \sqrt{2} \sin(\frac{3}{4}\pi+(s_1+s_2-2)\frac{\pi}{4})$,
leads to the well-known CHSH inequality:\cite{CHSH}
\begin{equation}
|E(1,1) + E(1,2) + E(2,1) - E(2,2) |\le 2.
\label{CHSH}
\end{equation}
The set of all $16$ inequalities (\ref{allbellineq}) is
equivalent to a {\it single} general Bell inequality:
\begin{equation}
 \sum_{s_1,s_2  =-1,1}  |
 \sum_{k_1,k_2 = 1,2} s^{k_1-1}_1 s^{k_2-1}_2  E({k_1},{k_2})|\leq 4.
\label{thebellineq}
\end{equation}
The equivalence of (\ref{thebellineq}) and (\ref{allbellineq}) is
evident, once one  recalls that, for real numbers $|a+
b|\leq c$ and $|a-b|\leq c$  if and only if $|a|+|b|\leq c$, and
writes down a generalization of this property to sequences of an
arbitrary length.

Whenever local realistic model exists inequality (\ref{thebellineq})
is satisfied by its predictions.
To prove the sufficiency of condition (\ref{thebellineq})
we construct a local realistic model
for any set of correlation functions which satisfy it,
i.e. we are interested in the local realistic models
$E^{LR}(k_1,k_2)$ such that they fully agree the measured correlations $E(k_1,k_2)$
for all possible observables $k_1,k_2=1,2$.
Recall that the set of local realistic correlation functions
can be put as (\ref{CORRELATION_POLYTOPE}),
with $\vec A = (A_1(\vec{n}_{1}^1), s_1 A_1(\vec{n}_{2}^1))$
and $\vec B = (A_2(\vec{n}_{1}^2), s_2 A_2(\vec{n}_{2}^2))$.
Let us ascribe for fixed $s_1,s_2$, a hidden probability that $A_j(\vec{n}_{1}^j) = s_j A_j(\vec{n}_{2}^j))$ 
(with $j=1,2$) in the form familiar from Eq. (\ref{thebellineq}):
\begin{equation}
P(s_1,s_2)=\frac{1}{4} |\sum_{k_2,k_2} s_1^{k_1-1} s_2^{k_2-1} E(k_1,k_2)|.
\label{PROB}
\end{equation}
Obviously these probabilities are positive.
However they sum up to identity only if inequality (\ref{thebellineq}) is saturated,
otherwise there is a ``probability deficit'', $\Delta P$.
This deficit can be compensated without affecting correlation functions.

First we construct the following structure, which is indeed the local
realistic model of the set of correlation functions if the
inequality is saturated:
\begin{equation}
\sum_{s_1,s_2=\pm 1} \Sigma(s_1,s_2) P(s_1,s_2) (1,s_1) \otimes (1,s_2),
\label{MODEL}
\end{equation}
where $\Sigma(s_1,s_2)$ is the sign of the expression within the modulus in Eq. (\ref{PROB}).
Now if $\Delta P > 0$ we add a ``tail'' to this expression given by:
\begin{equation}
\frac{\Delta P}{16} \sum_{\vec A, \vec B= (\pm 1,\pm 1)} \vec A \otimes \vec B.
\end{equation}
This ``tail'' does not contribute to the values of the correlation functions,
because it represents the fully random noise.
Thus, the sum of (\ref{MODEL}) is a valid local realistic model for 
$\hat E = (E(1,1),E(1,2),E(2,1),E(2,2))$.

Let us make some additional remarks.
In the four dimensional real space where both $\hat E_{LR}$ and $\hat E$
are defined one can find an orthonormal basis set $\hat S_{s_1s_2} = \frac{1}{2}(1,s_1)\otimes(1,s_2)$.
Within these definitions the hidden probabilities acquire a simple form:
\begin{equation}
P(s_1,s_2) = \frac{1}{2} |\hat S_{s_1s_2} \cdot \hat E|,
\end{equation}
where the dot denotes the scalar product in $\mathcal{R}^4$.
Now the local realistic models, $\hat E_{LR}$, can be expressed as:
\begin{equation}
\hat E_{LR} = \sum_{s_1,s_2=\pm1} 
|\hat S_{s_1s_2} \cdot \hat E| A_1(\vec{n}_{1}^1) A_2(\vec{n}_{1}^2) \hat S_{s_1s_2}.
\end{equation}
The modulus of any number $|x|$ can be split into $|x| = x \textrm{ sign}(x)$,
and we can always demand the product $A_1(\vec{n}_{1}^1) A_2(\vec{n}_{1}^2)$
to have the same sign as the expression inside the modulus.
Thus we have:
\begin{equation}
\hat E = \sum_{s_1,s_2=\pm1} 
(\hat S_{s_1s_2} \cdot \vec E) \hat S_{s_1s_2}.
\end{equation}
The expression in the bracket is the coefficient of tensor $\hat E$
in the basis $\hat S_{s_1s_2}$. These coefficients are then summed
over the same basis vectors, therefore 
the last equality appears.

In this way the set of inequalities (\ref{allbellineq}),
or its equivalent -- the single inequality (\ref{thebellineq}) --
is proven to be sufficient and necessary for the possibility of local realistic
description of correlation experiments on two qubits,
where both Alice and Bob measure one of two local settings.
This kind of reasoning can also be applied to arbitrary number of qubits.

\section{Many Qubits}

Exiting features of the GHZ states led to a rapid development of new Bell
inequalities for multiqubit systems.
Series of Bell inequalities for correlation functions were discovered by Mermin,
Ardehali, Belinskii and Klyshko \cite{MERMIN,ARDEHALI,BELINSKII}.
Below we present a derivation of series which form the complete
set of inequalities for $N$ parties, two-settings problem.

The generalisation of the approach presented for two-qubit case to many qubits is straightforward.
For $N$ particles the generalisation of identity (\ref{INEQ2})
consists of the sum of the products $A_j(\vec{n}_{1}^j) + s_j A_j(\vec{n}_{2}^j) = \pm 2$
for the $j$th party, and the summation is taken with
more general ``sign function'', $S(s_1,...,s_N)$, of $N$ parameters:
\begin{equation}
A_{12,...,12;S} \equiv \sum_{s_1,...,s_N =\pm1} 
S(s_1,...,s_N) \prod_{j=1}^N [ A_j(\vec{n}_{1}^j) + s_j A_j(\vec{n}_{2}^j)] =\pm 2^N, \label{INEQN}
\end{equation}
Since there are $2^{2^N}$ different sing functions $S$, the above formula leads to the set of $2^{2^N}$ Bell inequalities. 
Using the trick described above we can write a \emph{single} inequality
equivalent to set of all $2^{2^N}$ inequalities \cite{WW,WZ,ZB}:
\begin{equation}
 \sum_{s_1,...,s_N  =-1,1}|
 \sum_{k_1,...,k_N = 1,2} s^{k_1-1}_1 ... s^{k_N-1}_N  E({k_1},...,{k_N})|\leq 2^N.
\label{THEBELLINEQ_N}
\end{equation}
Many of these inequalities are trivial. e.g., if $S(s_1,...,s_N)=1$ for all arguments, we get the condition
$|E(1,1,...1)|\leq1$. Specific other choices give non-trivial inequalities. For example, for $N=3$, 
we can recover the inequality (equivalent to the GHZ argument) given by Mermin:\cite{MERMIN}
\begin{equation}
|E(2,1,1)+E(1,2,1)+E(1,1,2)-E(2,2,2)| \leq 2. \label{mermin}
\end{equation}

Up to now we have shown that if 
a local realistic model exists,
the general Bell inequality (\ref{THEBELLINEQ_N}) follows. The
converse is also true: whenever inequality (\ref{THEBELLINEQ_N})
holds one can construct a local realistic model for the
correlation function, in the case of a standard Bell experiment.
For $N$ particles the hidden probability that the predetermined outcomes 
of the $j$th observer are $A_j(\vec{n}_{1}^j) = s_j A_j(\vec{n}_{2}^j)$
is given by the form familiar from Eq. (\ref{THEBELLINEQ_N}):
\begin{equation}
P(s_1,...,s_N) = \frac{1}{2^N} \Big|\sum_{k_1,...,k_N=1,2} s_1^{k_1-1}...s_N^{k_N-1}E(k_1,...,k_N)\Big|. 
\end{equation}
The same steps as for two qubits above (now in the $\mathcal{R}^{2^N}$ space)
lead to the result that any correlation experiment satisfying (\ref{THEBELLINEQ_N})
can be explained within local realistic picture.
This establishes the general Bell inequality (\ref{THEBELLINEQ_N})
as a necessary and sufficient
condition for local realistic description of $N$ particle
correlation functions in standard Bell-type experiments.
That is one can claim that the set of Bell
inequalities represented by (\ref{THEBELLINEQ_N}) is complete.
This completeness implies that all series of Mermin $N$-qubit inequalities,
which give tight inequalities, are a subset of the inequalities generated
by (\ref{THEBELLINEQ_N}).
This also applies to tight Ardehali inequalities
and the full set of Belinskii-Klyshko inequalities \cite{ARDEHALI,BELINSKII}.

\section{More Than Two Settings}

A general way to establish a necessary and sufficient condition for local realistic description
is to define the facets of a correlation polytope \cite{PS,SLIWA,CG}.
However this is a computationally hard NP problem \cite{PITOWSKY}.
Even the complexity of printing out all Bell inequalities 
grows rapidly with the number of particles or settings involved! 
Now we present an efficient method for generation
of tight Bell's inequalities, which however do not form a complete set.
This method was invented by Wu and Zong \cite{WUZONG1,WUZONG2}, and generalized in Ref. \cite{LPZB}.

We start with the case of $N=3$ observers. Suppose that the first
two observers can choose between four settings, and the third one
between two settings. We denote such problem as $4 \times 4
\times 2$. 
We already know that the local realistic values satisfy the following algebraic
identity:
\begin{equation}
A_{12,12,S'} \equiv \sum_{s_1,s_2=\pm 1}
S'(s_1,s_2) [A_1(\vec n_1^1)+ s_1 A_1(\vec n_2^1)] 
[A_2(\vec n_1^2)+ s_2 A_2(\vec n_2^2)]=\pm 4, \label{kraj}
\end{equation}
where $S'(s_1,s_2)$ is any sign function, i.e. such that
$S'(s_1,s_2) = \pm 1$. 
In analogous way, we can define $A_{34,34,S''}$ by replacing
$A_1(\vec n_1^1),A_1(\vec n_2^1),A_2(\vec n_1^2),A_2(\vec n_2^2)$ by 
$A_1(\vec n_3^1),A_1(\vec n_4^1),A_2(\vec n_3^2),A_2(\vec n_4^2)$, respectively, and $S'$ by
$S''$. Depending on the value of $s= \pm 1$ one has $(A_{12,S'} +
s A_{34,S''})=\pm 8,$ or $0$. By analogy to (\ref{kraj}) one has:
\begin{equation}
A_{1234,12} \equiv
\sum_{s_1,s_2=\pm 1} \hspace{-2mm}  
S(s_1,s_2) [A_{12,S'}+s_1 A_{34,S''}][A_3(\vec n_1^3)+ s_2 A_3(\vec n_2^3)]=\pm 16.
\label{GEN}
\end{equation}
After averaging over many runs of the experiment, and introducing
the correlation functions $E(i,j,k) \equiv \langle A_1(\vec n_i^1) A_2(\vec n_j^2) A_3(\vec n_k^3)
\rangle_{\textrm{avg}}$ one obtains multisetting Bell's inequalities.
Because of the freedom to choose the sign functions $S, S', S''$,
we have $(2^4)^3 = 2^{12}$ Bell's inequalities.

All these inequalities can be reduced to single ``generating'' inequality,
in which all the sign functions $S, S', S''$
are non-factorable.
The choice of factorable sign function
is equivalent to having a non-factorable one, and some of the
local measurement settings equal \cite{LPZB}.
The ``generating" Bell's inequality can be chosen as
(here all sign functions are equal to $\sqrt{2}\sin(\frac{3}{4}\pi + (s_1+s_2-2)\frac{\pi}{4})$):
\begin{eqnarray}
&& \Big| \Big \langle 
\Big[ A_3(\vec n_1^3) + A_3(\vec n_2^3)\Big] 
\Big[ A_1(\vec n_1^1)[(A_2(\vec n_1^2) + A_2(\vec n_2^2)] + A_1(\vec n_2^1)[A_2(\vec n_1^2) - A_2(\vec n_2^2)] \Big] 
\label{442_GEN_INEQ} \\
&& + \Big[ A_3(\vec n_1^3) - A_3(\vec n_2^3) \Big] 
\Big[ A_1(\vec n_3^1)[A_2(\vec n_3^2) + A_2(\vec n_4^2)] + A_1(\vec n_4^1) [A_2(\vec n_3^2) - A_2(\vec n_4^2)] \Big] 
\Big\rangle_{\textrm{avg}} \Big| \nonumber \\
&& \le 4. \nonumber
\end{eqnarray}
All other inequalities can be obtained by changes $A_j(\vec n_k^j) \to -A_j(\vec n_k^j)$.

The method can be generalized to various choices of the number of
parties and the measurement settings. Now we present the
$2^{N-1} \times 2^{N-1} \times 2^{N-2}\times ... \times 2$ case. 
Consider $N=4$ observers. 
We start with the identity (\ref{GEN}). One can introduce a similar
formula for the settings $\{5,6,7,8\}$, for the first two
observers, and $\{3,4\}$, for the third one. The fourth observer
chooses between two settings with local realistic values $A_4(\vec n_1^4)$ and $A_4(\vec n_2^4)$.
Applying the same method as before, one obtains an identity which
generates Bell's inequalities of the $8 \times 8 \times 4 \times
2$ type:
\begin{eqnarray}
&&\sum_{s_1,s_2 = \pm 1} S(s_1,s_2) 
[A_{1234,12}+s_1 A_{5678,34}] [A_4 (\vec n_1^4) + s_2 A_4 (\vec n_2^4)] = \pm 64,
\label{8842IDENTITY}
\end{eqnarray}
where $A_{1234,12}$ and  $A_{5678,34}$ depend on some three sign
functions. One may apply this method iteratively,
increasing the number of observers by one, to obtain inequalities
involving exponential (in $N$) number of measurement settings.

As another example we construct the inequalities involving $N$ partners,
where first $N-1$ observers choose one of $4$ settings and the last one
chooses between $2$ settings.
We use the local realistic quantity $A_{12,...,12}$ defined in Eq. (\ref{INEQN})
for $N-1$ parties choosing between $2$ settings each:
\begin{equation}
A_{12,...,12} \equiv 
\sum_{s_1,...,s_{N-1} =\pm1} 
S'(s_1,...,s_{N-1}) 
\prod_{j=1}^{N-1} [ A_j(\vec{n}_{1}^j) + s_j A_j(\vec{n}_{2}^j)] =\pm 2^{N-1},
\label{121212}
\end{equation}
and analogically introduce $A_{34,...,34}$ for another pair of observables
available to each party. The sign function in $A_{34,...,34}$
can be different from $S'(s_1,...,s_{N-1})$ in Eq. (\ref{121212}).
By including the $N$th observer, who can choose between $2$ measurement settings,
we obtain:
\begin{equation}
\sum_{s_1,s_1 =\pm1} 
S(s_1,s_2)
(A_{12,...,12} + s_1 A_{34,...,34})(A_N(\vec{n}_{1}^N) + s_2 A_N(\vec{n}_{2}^N)) = \pm 2^{N+1}.
\end{equation}
One can use this expression for generating Bell inequalities
for $N$ observers in the same way as it was previously done.

In order to show the full strength  of the method our 
next example gives a family of  Bell inequalities 
for $N=5$ qubits, which involves eight settings for first two
observers and four settings for the other three.
We take the identity $A_{1234,12}$ defined in (\ref{GEN}), 
valid for the $4 \times 4 \times 2$ case of three observers,
and define a similar quantity for another set of $4 \times 4 \times 2$ observables,
namely $A_{5678,34}$.
Note that the sign functions entering $A_{5678,34}$
can be different from those entering $A_{1234,12}$.
For the other two observers we introduce:
\begin{equation}
A_{12,12} \equiv \sum_{s_1,s_2 = \pm 1}S'(s_1,s_2)(D_1+s_1D_2)(E_1+s_2E_2)=\pm 4
\end{equation}
and a similar expression for another pair of observables $D_3, D_4$ and $E_3, E_4$:
\begin{equation}
A_{34,34}= \sum_{s_1,s_2 = \pm 1}S''(s_1,s_2)(D_3+s_1D_4)(E_3+s_2E_4)=\pm 4.
\end{equation}
In the next step we get the following algebraic identity which
can be used, via averaging, to generate a family of $2^{28}$ Bell
inequalities:
\begin{equation}
\sum_{s_1, s_2 = \pm 1}S(s_1,s_2)
(A_{1234,12}+s_1 A_{5678,34})
(A_{12,12} + s_2 A_{34,34}) = \pm 256.
\end{equation}

It is clear that there is no bound in extending this type of derivations. 
Finally let us note that all the inequalities with lower number of settings 
can be obtained from our construction by making some
of the local settings identical.

The multisetting inequalities constructed by the above
procedure are tight. Consider the case of $4 \times 4 \times 2$ inequalities.
The left hand side of the identity
(\ref{GEN}) is equal to $\pm 16$ for any combination of
predetermined local realistic results. In a 32 dimensional real space, one can
build a convex polytope, containing all possible local realistic models of the
correlation functions for the specified settings, with vertices
 given by the tensor products of 
$\hat v=(A_1(\vec n_1^1),A_1(\vec n_2^1),A_1(\vec n_3^1),A_1(\vec n_4^1))
\otimes(A_2(\vec n_1^2),A_2(\vec n_2^2),A_2(\vec n_3^2),A_2(\vec n_4^2))
\otimes(A_3(\vec n_1^3), A_3(\vec n_2^3))$. 
It has $256$ different vertices. Tight Bell inequalities define the
half-spaces in which is the polytope,  which contain a face of it
in their border hyperplane. If 32 linearly independent vertices
belong to a hyperplane, this hyperplane defines a tight
inequality. Half of the vertices give in (\ref{GEN}) the value
$16$ and  another half give $-16$.
Every vertex, $\hat v$ from the first set has a
partner $- \hat v$ in the second one. 
Next notice that 
any set of $128$
vertices $\hat v$, which does not contain pairs $\hat v$ and $- \hat v$ contains
a set of 32 linearly independent points (basis). Thus, each
inequality is tight. This reasoning can be adapted to all
inequalities discussed here.

Most importantly, the multisetting inequalities reveal violation
of local realism of classes of states, for which standard inequalities,
with two measurement settings per side, are satisfied.

\section{Quantum Violations}

Bell inequalities are interesting only if there exists a certain state which violates them
(for certain measurement settings).
Let us derive the condition for violation of
the general Bell inequality involving two measurement settings, (\ref{THEBELLINEQ_N}),
by an arbitrary two-qubit state. 
This result is a generalisation and reformulation of the condition
given for two qubits by the Horodecki Family \cite{HORODECKI}.
We will use a decomposition of general mixed state of
$N$ qubits in terms of the identity operator $\sigma_0= I$ 
 and the Pauli operators
$\sigma_i$ for three orthogonal directions $i\in \{1,2,3\}$, given 
by:
\begin{equation}
 \rho=\frac{1}{2^N} \sum_{k_1,...,k_N=0}^{3}
T_{k_1...k_N} \sigma_{k_1} \otimes ...\otimes \sigma_{k_N},
\end{equation}
where all the $\sigma_i$ operators act in the Hilbert space of individual qubits.
 The
(real) coefficients $T_{k_1...k_N}$, with $k_j=1,2,3$, form the 
correlation tensor $\hat{T}$.
Note further that $T_{k_1...k_N} = Tr(\rho \sigma_{k_1} \otimes ...\otimes \sigma_{k_N})$,
i.e. all coefficients are directly experimentally accessible.

The full set of inequalities for the $2 \times 2$ problem is
derivable from the identity (\ref{kraj}) where we put non-factorable sign function:
\begin{equation}
\Big| \Big\langle 
(A_1+ A_2)B_1 + (A_1 - A_2)B_2 \Big\rangle_{\textrm{avg}} \Big| \le 2. \label{full22}
\end{equation}
All other inequalities are obtainable by all possible sign changes
$X_k \to -X_k$ (with $k=1,2$ and $X = A,B$).
The quantum correlation function  $E(\vec a_k,\vec b_l)$ is given by
the scalar product of the correlation tensor $\hat T$ with the
tensor product of the local measurement settings represented by
unit vectors $\vec a_k \otimes \vec b_l$, i.e. $E(\vec a_k,\vec b_l) = (\vec a_k \otimes \vec b_l) \cdot \hat T$. 
Thus, the condition for a
quantum state endowed with the correlation tensor $\hat T$ to
satisfy the inequality (\ref{full22}), is that for all directions
$ \vec a_1, \vec a_2, \vec b_1, \vec b_2$ one has
\begin{equation}
\Big| \Big[ (\frac{\vec a_1 + \vec a_2}{2}) \otimes \vec b_1
+ (\frac{\vec a_1 - \vec a_2}{2}) \otimes \vec b_2 \Big] \cdot \hat T \Big| \le 1,
\label{QUANTUM_INEQ}
\end{equation}
where both sides of (\ref{full22}) were divided by $2$.

Note that the pairs of local vectors define the
``local measurement planes''.
Here we shall find the conditions for (\ref{QUANTUM_INEQ}) to hold for two,
arbitrary but fixed, measurement planes, one for each observer.
Therefore only those components of $\hat T$ are relevant which describe
measurements in these two planes.
Thus $\hat T$ is effectively described by a $2 \times 2$ matrix, or tensor $\hat T'$.

Notice that $\vec A_{\pm} = \frac{1}{2}(\vec a_1 \pm \vec a_2)$
satisfy the following relations: $\vec A_+ \cdot \vec A_- = 0$ and 
$||\vec A_+ ||^2 + ||\vec A_- ||^2 = 1$.
Thus $\vec A_+ + \vec A_-$ is a unit vector,
and $\vec A_\pm$ represent its decomposition into two orthogonal vectors.
Thus if one introduces unit vectors $\vec a_\pm$ such that
$\vec A_\pm = a_\pm \vec a_\pm$, one has $a_+^2 + a_-^2 = 1$.
Thus one can put (\ref{QUANTUM_INEQ}) into the following form:
\begin{equation}
|\hat S \cdot \hat T'| \le 1,
\end{equation}
where $\hat S = a_+ \vec a_+ \otimes \vec b_1 + a_- \vec a_- \otimes \vec b_2$.
Since $\vec a_+ \cdot \vec a_- = 0$,
one has $\hat S \cdot \hat S = 1$, i.e. $\hat S$ is a tensor of unit norm.
Any tensor of unit norm, $\hat U$, has the following Schmidt decomposition
$\hat U = \lambda_1 \vec v_1 \otimes \vec w_1 + \lambda_2 \vec v_2 \otimes \vec w_2$,
where $\vec v_i \cdot \vec v_j = \delta_{ij}, \vec w_i \cdot \vec w_j = \delta_{ij}$
and $\lambda_1^2 + \lambda_2^2 = 1$.
The (complete) freedom of  the choice of the measurement directions
$\vec b_1$ and $\vec b_2$, allows one, 
by choosing $\vec b_2$ orthogonal to $\vec b_1$,
to put $\hat S$ in the form isomorphic with $\hat U$.
The freedom of choice of $\vec a_1$ and $\vec a_2$ allows
$\vec A_+$ and $\vec A_-$ to be arbitrary orthogonal unit vectors,
and $\vec a_+$ and $\vec a_-$ to be also arbitrary.
Thus $\hat S$ can be equal to any unit tensor.
Therefore to get the maximum of the left hand side of (\ref{QUANTUM_INEQ})
we put $\hat S = \frac{1}{||\hat T'||} \hat T'$,
and the maximum is $||\hat T'|| = \sqrt{\hat T' \cdot \hat T'}$.
Thus,
\begin{equation}
\max\big[\sum_{k,l=1,2} T_{kl}^2 \big] \le 1,
\label{t2-22}
\end{equation}
where 
the maximization
is taken over all local coordinate systems of two observers,
is the necessary and sufficient condition for the inequality (\ref{THEBELLINEQ_N})
to hold for quantum predictions.

For $2^{N-1} \times 2^{N-1} \times ... \times 2$ multisetting inequalities 
we can derive similar simple conditions for larger number of qubits. 
Consider the case of three qubits. For this situation we have the following inequality:
\begin{equation}
\Big| \Big\langle  \sum_{s_1,s_2} S(s_1,s_2)
[A_{12,S'}+s_1 A_{34,S''}] [C_1+s_2 C_2]
\Big\rangle_{\textrm{avg}} \Big| \le 16,
\label{442_INEQ}
\end{equation}
where $S,S',S''$ are some non-factorable sign functions.
The three-qubit quantum correlation functions 
$E(\vec a_i, \vec b_j, \vec c_k)$  can be represented
as  $(\vec a_i \otimes \vec c_j
\otimes \vec c_k) \cdot \hat T$ (with the same meaning of the
symbols as before; $\hat T$ is now a three index tensor). 
Thus the
condition for the $4 \times 4 \times 2$ inequalities to hold, in
the quantum case, transforms into
\begin{equation}
|[\hat A_{12,S'} \otimes (\vec c_1 + \vec c_2) + \hat A_{34,S''}
\otimes (\vec c_1 - \vec c_2)] \cdot \hat T |\le 8, \label{KO}
\end{equation}
where e.g. \begin{equation}
\hat A_{12,S'} =\sum_{s_1,s_2= \pm 1} S'(s_1,s_2) (\vec a_1 + s_1 \vec a_2) \otimes
(\vec b_1 + s_2 \vec b_2). \nonumber
\end{equation}
To write down (\ref{KO}) we have used the freedom of introducing the sign changes 
$\vec X_i \to - \vec X_i$, compare (\ref{442_GEN_INEQ}).
By defining $\frac{1}{2}(\vec c_1 \pm \vec c_2) = c_\pm \vec c_\pm$
which have the similar properties as $a_{\pm}$ and $\vec a_{\pm}$, inequality (\ref{KO})
transforms to:
\begin{equation}
|c_+ \hat A_{12,S'} \otimes \vec c_+ \cdot \hat T + c_- \hat
A_{34,S''} \otimes \vec c_- \cdot \hat T |\le 4.
\end{equation}
One can always choose $c_+$ and $c_-$ that maximize the left 
hand side. Since $c_+^2 + c_-^2 =1$ this leads to the condition:
\begin{equation}
[\hat A_{12,S'} \cdot \hat T^{(+)}]^{2} + [\hat A_{34,S''} \cdot
\hat T^{(-)}]^{2} \leq 4^2,
\end{equation}
where $\hat T^{(\pm)}$ is
defined by $T^{(\pm)}_{ij}=\sum_{k=1}^3 (c_\pm)_k T_{ijk}$, where in turn
$(c_\pm)_k$ is the $k$-th component of vector $\vec c_\pm$. Note that 
since $\vec c_+$ and $\vec c_-$ are orthogonal and normalized
this procedure amounts to fixing of two (new) Cartesian axes for 
the third observer, and accordingly transforming the correlation tensor.
Since
$\hat A_{12,S'}$ depends on different vectors than $\hat
A_{34,S''}$, one can maximize the two terms separately. 
Furthermore, since the problem of maximization of
$\hat A_{nm,S} \cdot \hat T^{(\pm)}$ is equivalent to the $2 \times
2$ case studied earlier, the overall maximization process gives
the following \emph{necessary and sufficient} condition for
quantum correlations to satisfy the inequality (\ref{442_INEQ}):
\begin{equation}
\max \sum_{x=1,2} \sum_{k_x,l_x=1,2} T_{k_x l_x x}^2 \le 1.
\label{NS_CONDITION_MULTI}
\end{equation}
When compared with the \emph{sufficient} condition for
$2 \times 2 \times 2$ inequalities to hold, namely: \cite{ZB}
\begin{equation}
\max \big[\sum_{k,l,m=1,2} T_{klm}^2\big] \le 1,
\label{S_COND_THREE}
\end{equation}
the new 
condition is {\em more demanding} because the
Cartesian coordinate systems denoted by the indices $k_1,l_1$ and
$k_2,l_2$ do not have to be the same.
Note further that if one uses in both terms in (\ref{NS_CONDITION_MULTI})
\emph{the same planes of observations} (for the first two observers)
then the condition reduces to (\ref{S_COND_THREE}),
which now is a sufficient and necessary one (but for $4 \times 4 \times 2$ settings).
Thus the condition is necessary and sufficient whenever the vectors
defining measurements for each party are limited to one plane.

In a similar way one can reach analogous conditions for
violation of $2^{N-1} \times 2^{N-1} \times 2^{N-2} \times ...
\times 2$ inequalities by quantum predictions.
In the Table 1 we present these conditions for small $N$.
\begin{table}
\caption{The examples of necessary and sufficient conditions for violation of
multisetting inequalities.}
\begin{center}
\begin{tabular}{c|c|c}
\hline  \hline
$N$ & case & $C_N$  (the condition) \\ \hline \hline
$2$ & $2 \times 2$ & $\sum_{k,l=1,2} T_{kl}^2 \le 1$ \\ \hline
$3$ & $4 \times 4 \times 2$ & $\sum_{k,l=1,2} T_{kl2}^2 + \sum_{k',l'=1,2} T_{k'l'1}^2 \le 1$  \\ \hline
$4$ & $8 \times 8 \times 4 \times 2$ & 
$\sum_{k_1,l_1=1,2} T_{k_1l_122}^2 + \sum_{k_2,l_2=1,2} T_{k_2l_212}^2 +$ \\
&&
$\sum_{k_3,l_3=1,2} T_{k_3l_321}^2 + \sum_{k_4,l_4=1,2} T_{k_4l_411}^2 \le 1 $
\\ \hline 
\end{tabular}
\end{center}
\end{table}
One can note a useful recurrence that can be used to write down the condition for arbitrary $N$. 
Let us define:
\begin{equation}
\mathcal{C}_2 \equiv \sum_{k,l=1,2} T_{kl}^2.
\end{equation}
Then the condition for two qubits reads: $\max (\mathcal{C}_2) \le 1$.
Next let us put a recursive definition:
\begin{equation}
\mathcal{C}_N = [\mathcal{C}_{N-1}]_{\oplus 2} + [\mathcal{C}_{N-1}]_{\oplus 1}',
\end{equation}
where $[\mathcal{C}_{N-1}]_{\oplus k}$ is the expression in the condition for $N-1$ qubits
in which the correlation tensor elements $T_{i_1...i_{N-1}}$
are replaced by $T_{i_1...i_{N-1}k}$, i.e. elements of the $N$-qubit correlation tensor.
The ``prime'' denotes the fact that the second term does not involve components
of $\hat T$ in the same set of coordinate systems 
(for the first $N-1$ observers) as the unprimed term.

The sufficient and necessary condition for $N$ qubits to satisfy
all $2^{N-1} \times 2^{N-1} \times 2^{N-2} \times ... \times 2$
inequalities, within this convention reads:
\begin{equation}
\max(\mathcal{C}_N) \le 1.
\end{equation}

\section{Gisin's Problem}

The theorem of Gisin states that {\it any} pure non-product state violates local realism, 
i.e.
there are sets of measurements that can be performed on the state
which cannot be described within local realistic picture \cite{GISIN,GISINPERES}.
This theorem formalizes the intuition that entanglement is a purely quantum phenomenon.
Using the approach presented here we can write down the following proof of Gisin's theorem for \emph{two qubits}. 
Any state of two qubits is given in its Schmidt basis by $|\psi \rangle = 
\cos{\alpha} |00\rangle + \sin{\alpha} |11\rangle$, with $\alpha \in [0,\pi/4]$. 
The correlation tensor of this state has the following coefficients:
$T_{xx} = \sin{2\alpha}$, $T_{yy} =  - \sin{2\alpha}$, $T_{zz} = 1$. 
Therefore the necessary and sufficient condition for local realism is violated
for all non-product ($\alpha \neq 0$) states:
\begin{equation}
\sum_{k,l=\{x,z\}} T_{kl}^2 = 1 + \sin^2{2\alpha} >1.
\end{equation}

As we shall show here the series of the two-settings inequalities (\ref{THEBELLINEQ_N})
for $N$ qubits are failing to show violation of local realism for an important
class of entangled pure states.
This looks, superficially, as a counterexample to Gisin's theorem.
But the theorem uses all possible measurement scenarios, which is not the case
in the case of inequalities (\ref{THEBELLINEQ_N}).
Further, as it will be shown, the more-than-two setting inequalities
do show violations for this class of states.
This is a strong argument for continuation of the research
towards finding inequalities with even more general structure.

Scarani and Gisin noticed a surprising feature of
the following state:\cite{SCARANI}
\begin{equation}
|\psi\rangle = \cos\alpha|0...0\rangle+\sin\alpha|1...1\rangle. \label{STATE}
\end{equation}
They showed that for $\sin{2\alpha}\! \leq \!
1/\sqrt{2^{N-1}}$ the states (\ref{STATE}) do not violate
the Mermin-Ardehali-Belinskii-Klyshko (MABK) inequalities \cite{MERMIN,ARDEHALI,BELINSKII}.
 This has been numerically obtained for
$N\!=\!3,4,5$ and conjectured for $N\!>\!5$. Their result
contrasts the case of two qubits and is highly counterintuitive as
the states (\ref{STATE}) are generalization of the GHZ states,
which violate maximally the MABK inequalities \cite{GHZ}. 
The problem of Scarani and Gisin was later analysed
using the general two-qubit correlation function Bell inequality 
for arbitrary number of qubits (\ref{THEBELLINEQ_N}) \cite{ZBLW}.
The family (\ref{STATE})
does not violate {\it any} such Bell inequality 
if $\alpha$ and $N$ are such that $\sin2\alpha\leq 1/\sqrt{2^{N-1}}$ 
and \emph{$N$ is odd}.
However for even number of qubits the general 
Bell inequality imposes stronger constraints
for local realistic description, than MABK inequalities. 

What are the reasons for the completely different behaviour for $N$
even and $N$ odd? The expression $\sum_{k_1,...,k_N=x,y} T_{k_1...k_N}^2$,
which appears in the condition for violation of two-setting inequalities,
can be understood as a ``total measure of the strength
of correlations'' in mutually complementary sets of local
measurements (as defined by the summation over $x$ and $y$) \cite{ESSENCE}. 
Then the unity on the right-hand side of
the condition is the classical limit for the amount of correlations.
Specifically, pure product states cannot exceed the limit of 1, as
they can show perfect correlations in one set of local
measurement directions only.
In contrast, entangled states can
show perfect correlations for more than one such set \cite{ESSENCE}.
Now, only if $N$ is even, the states (\ref{STATE})
already show perfect correlation between measurements along
$z$-directions (as the product is then always +1) reaching
therefore the classical limit
(we assume $\sigma_z$ eigenstates give the qubit computational basis). 
Yet, they also show additional
correlations in other, complementary, directions. In the case of
$N$ odd, however, there are no perfect correlations 
along $z$-direction and the correlations in the
complementary directions do not suffice to violate the bound of $1$.

Since the condition for multipartite two-setting inequalities to hold:
\begin{equation}
\max \sum_{k_1,...,k_N=1,2} T_{k_1...k_N}^2 \le 1,
\end{equation}
is only necessary, if a state satisfies it one cannot draw any conclusions
about its possible local realistic model.
Thus it could be that the states (\ref{STATE}) can violate local realism
if we use some more optimal tool to test it.
Indeed multisetting inequalities described extend the class of
quantum states which do not admit local realistic explanation.
Also the generalized GHZ states cannot be classically explained.
Their nonvanishing correlation tensor components are
(directions $1,2,3$
are denoted by $x,y,z$; the basis $\{|0\rangle,|1\rangle\}$ is
the eigenbasis of $\sigma_z$): $T_{z...z} = \cos{2\alpha}$, for
$N$ odd, and 1 for $N$ even, $T_{x...x} = \sin{2\alpha}$, and the
components with $2k$ indices equal to $y$ and the rest equal to
$x$ take the value $(-1)^k \sin{2\alpha}$ (there are $2^{N-2}$
such components). Let us assume that the last observer can choose only between
settings $x$ and $z$. Thus, we obtain for the condition for
violation of multisetting Bell's inequality for \emph{odd} number of observers
(generalization of the condition (\ref{NS_CONDITION_MULTI})) 
\begin{eqnarray}
&\sum_{k_1,...,k_{N-1}=x,y} T^2_{k_1...k_{N-1}x}+ \sum_{k_1,
..., k_{N-1} = x,z}
T^2_{k_1...k_{N-1}z}&\nonumber \\ &= 2^{N-2} \sin^2{2\alpha} +
\cos^2{2\alpha} > 1.& 
\end{eqnarray}
Thus,  the Bell's inequalities are violated for the whole range of $\pi/4\geq\alpha>0$ and for
arbitrary $N$ in contrast to the case of standard Bell's
inequalities.

\section{Prospects}

We have described some series of Bell's inequalities for arbitrary number of qubits
involving $N$-particle correlation functions.
In the case of two settings per observer the inequalities form a complete set,
which gives the necessary and sufficient condition for local realistic description
of correlation experiments.
Even for this simple scenario
it is still an open question how would the necessary and sufficient condition look like
for more general case involving all correlations of all ranks between observers,
such as those described by correlation functions for $N-1$ parties, etc.
(generalisations of CH inequality).
The other problem is to consider more settings per observer.
A step towards this direction was recently made by one of us,
who constructed a full set of inequalities for three settings per observer
and arbitrary number of parties \cite{NEWZUK}.
The final aim is to construct such inequalities for arbitrary
experimental setup.

It is clear that only entangled states can violate Bell's inequalities,
as any separable state has a local realistic model.
For pure entangled states there always exists Bell's inequality that
is violated by certain measurements performed on the state.
But which mixed states violate local realism? Perhaps
a step forward could be achieved by introduction of other,
additional to Bell's (local realism) assumptions, based on the fundamental
symmetries of physical laws, which would narrow the class of physically admissible local realistic theories \cite{ROTINV,NEWLZ}.

The work is part of the MNiI Grant no. 1 P03 04927. MZ acknowledges Professorial Subsidy of FNP.
WL and TP acknowledges a FNP stipend.

\end{document}